# Performance of the First ANTARES Detector Line

M. Ageron[a], J.A. Aguilar[b], A. Albert[c], F. Ameli[d], M. Anghinolfi[e], G. Anton[f], S. Anvar[g], M. Ardid[h], J-J. Aubert[a], J. Aublin[i], R. Auer[f], S. Basa[j], M. Bazzotti[k], Y. Becherini[i,1], V. Bertin[a], S. Biagi[k], A. Bigi[m], C. Bigongiari[b], M. Bou-Cabo[h], M.C. Bouwhuis[n], R. Bruijn[n,o], J. Brunner[a], G.F. Burgio[p], J. Busto[a], F. Camarena[h], A. Capone[d], G. Carminati[k], J. Carr[a], D. Castel[c], E. Castorina[m], V. Cavasinni[m], S. Cecchini[k,q], Ph. Charvis[r], T. Chiarusi[k], M. Circella[s], C. Colnard[n], R. Coniglione[t], H. Costantini[e], N. Cottini[l], P. Coyle[a], G. De Bonis[d], P. Decowski[n], I. Dekeyser[u], A. Deschamps[r], C. Donzaud[i,v], D. Dornic[c], D. Drouhin[c], F. Druillole[g], T. Eberl[f], J-P. Ernenwein[c], S. Escoffier[a], E. Falchini[m], F. Fehr[f], V. Flaminio[m], K. Fratini[e], J-L. Fuda[u], G. Giacomelli[k], K. Graf[f], G. Guillard[w], G. Hallewell[a], Y. Hello[r], J.J. Hernández-Rey[b], J. Höβl[f], M. de Jong[n], N. Kalantar-Nayestanaki[x], O. Kalekin[f], A. Kappes[f], U. Katz[f], P. Kooijman[y,o,n], C. Kopper[f], A. Kouchner[i], W. Kretschmer[f], S. Kuch[f], R. Lahmann[f], P. Lamare[g], G. Lambard[a], H. Laschinsky[f], J. Lavalle[a], H. Le Provost[g], D. Lefèvre[u], G. Lelaizant[a], G. Lim[n,o], D. Lo Presti[p], H. Loehner[x], S. Loucatos[i], F. Louis[g], F. Lucarelli[d], K. Lyons[w], S. Mangano[n], M. Marcelin[j], A. Margiotta[k], J.A. Martinez-Mora[h], G. Maurin[i], A. Mazure[j], M. Melissas[a], E. Migneco[t], T. Montaruli[s,1], M. Morganti[m], L. Moscoso[i,l], H. Motz[f], C. Naumann[f,l], R. Ostasch[f], G.E. Păvălaş[z], P. Payre[a], J. Petrovic[n], C. Petta[p], P. Piattelli[t], C. Picq[l], R. Pillet[r], V. Popa[z], T. Pradier[w], E. Presani[n], C. Racca[w], A. Radu[z], C. Reed[a], C. Richardt[f], M. Rujoiu[z], M. Ruppi[s], G.V. Russo[p], F. Salesa[b], P. Sapienza[t], F. Schoeck[f], J-P. Schuller[l], R. Shanidze[f], F. Simeone[d], M. Spurio[k], G. van der Steenhoven[n], C. Tamburini[u], L. Tasca[j], S. Toscano[b], M. Vecchi[d], P. Vernin[l], G. Wijnker[n], E. de Wolf[o,n], D. Zaborov[α], J.D. Zornoza[b], J. Zúñiga[b]

[a] CPPM - Centre de Physique des Particules de Marseille, CNRS/IN2P3 et Université de la Méditerranée, 163 Avenue de Luminy, Case 902, 13288 Marseille Cedex 9, France

[b] IFIC - Instituto de Física Corpuscular, Edificios Investigación de Paterna, CSIC - Universitat de València, Apdo. de Correos 22085, 46071 Valencia, Spain

[c] GRPHE - Groupe de Recherche en Physique des Hautes Energies, Université de Haute Alsace, 61 Rue Albert Camus, 68093 Mulhouse Cedex, France

[d] Dipartimento di Fisica dell'Università "La Sapienza" e Sezione INFN, P.le Aldo Moro 2, 00185 Roma, Italy

[e] Dipartimento di Fisica dell'Università e Sezione INFN, Via Dodecaneso 33, 16146 Genova, Italy

[f] Friedrich-Alexander-Universität Erlangen-Nürnberg, Physikalisches Institut, Erwin-Rommel-Str. 1, D-91058 Erlangen, Germany

[g] Direction des Sciences de la Matière - Institut de Recherche sur les lois Fondamentales de l'Univers - Service d'Electronique des Détecteurs et d'Informatique, CEA Saclay, 91191 Gif-sur-Yvette Cedex, France

[h] Institut de Gestió Integrada de Zones Costaneres (IGIC) - Universitat Politècnica de Valéncia. Cra. Nazaret-Oliva S/N E-46730 Gandia, València, Spain

[i] APC - Laboratoire AstroParticule et Cosmologie, UMR 7164 (CNRS, Université Paris 7 Diderot, CEA, Observatoire de Paris) 10, rue Alice Domon et Léonie Duquet 75205 Paris Cedex 13, France

[j] LAM - Laboratoire d'Astrophysique de Marseille, CNRS/INSU et Université de Provence, Traverse du Siphon - Les Trois Lucs, BP 8, 13012 Marseille Cedex 12, France

[k] Dipartimento di Fisica dell'Università e Sezione INFN, Viale Berti Pichat 6/2, 40127 Bologna, Italy

[l] Direction des Sciences de la Matière - Institut de Recherche sur les lois Fondamentales de l'Univers - Service de Physique des Particules, CEA Saclay, 91191 Gif-sur-Yvette Cedex, France

[m] Dipartimento di Fisica dell'Università e Sezione INFN, Largo B. Pontecorvo 3, 56127 Pisa, Italy

[n] FOM Instituut voor Subatomaire Fysica Nikhef, Kruislaan 409, 1098 SJ Amsterdam, The Netherlands

[o] Universiteit van Amsterdam, Instituut voor Hoge-Energiefysica, Kruislaan 409, 1098 SJ Amsterdam, The Netherlands

[p] Dipartimento di Fisica ed Astronomia dell'Università e Sezione INFN, Viale Andrea Doria 6, 95125 Catania, Italy

[q] INAF-IASF, via P. Gobetti 101, 40129 Bologna, Italy

[r] GéoSciences Azur, CNRS/INSU, IRD, Université de Nice Sophia-Antipolis, Université Pierre et Marie Curie - Observatoire Océanologique de Villefranche, BP48, 2 quai de la Darse, 06235 Villefranche-sur-Mer Cedex, France

[s] Dipartimento Interateneo di Fisica e Sezione INFN, Via E. Orabona 4, 70126 Bari, Italy

[t] INFN - Laboratori Nazionali del Sud (LNS), Via S. Sofia 44, 95123 Catania, Italy

[u] COM - Centre d'Océanologie de Marseille, CNRS/INSU et Université de la Méditerranée, 163 Avenue de Luminy, Case 901, 13288 Marseille Cedex 9, France

[v] Université Paris-Sud 11 - Département de Physique - F - 91403 Orsay Cedex, France

[w] IPHC-Institut Pluridisciplinaire Hubert Curien, Université Louis Pasteur (Strasbourg 1) et IN2P3/CNRS, 23 rue du Loess, BP 28, 67037 Strasbourg Cedex 2, France

[x] Kernfysisch Versneller Instituut (KVI), University of Groningen, Zernikelaan 25, 9747 AA Groningen, The Netherlands

[y] Universiteit Utrecht, Faculteit Betawetenschappen, Princetonplein 5, 3584 CC Utrecht, The Netherlands

[z] Institute for Space Sciences, R-77125 Bucharest, Măgurele, Romania

[α] ITEP - Institute for Theoretical and Experimental Physics, B. Cheremushkinskaya 25, 117259 Moscow, Russia

---

[1] On leave at University of Wisconsin -- Madison, 53715, WI, USA

# Abstract

In this paper we report on the data recorded with the first Antares detector line. The line was deployed on the 14[th] of February 2006 and was connected to the readout two weeks later. Environmental data for one and a half years of running are shown. Measurements of atmospheric muons from data taken from selected runs during the first six months of operation are presented. Performance figures in terms of time residuals and angular resolution are given. Finally the angular distribution of atmospheric muons is presented and from this the depth profile of the muon intensity is derived.

## 1. Introduction

Ever since Markov discussed the possibility of detecting cosmic neutrinos using the Cherenkov effect in sea water [1], many experimental groups have attempted to perform such an experiment. The DUMAND collaboration pioneered the technique in an experiment off the coast of Hawaii [2]. This was followed by the Baikal collaboration who built the first full detector in the fresh water of Lake Baikal [3]. Since then the AMANDA/IceCube collaboration has installed and operated large neutrino Cherenkov detectors in the ice of Antarctica [4]. More recently several collaborations have begun to investigate the Mediterranean Sea as a place to install a neutrino detector. The NESTOR [5] and NEMO [6] collaborations have deployed and operated prototypes of their detectors for short periods of time.

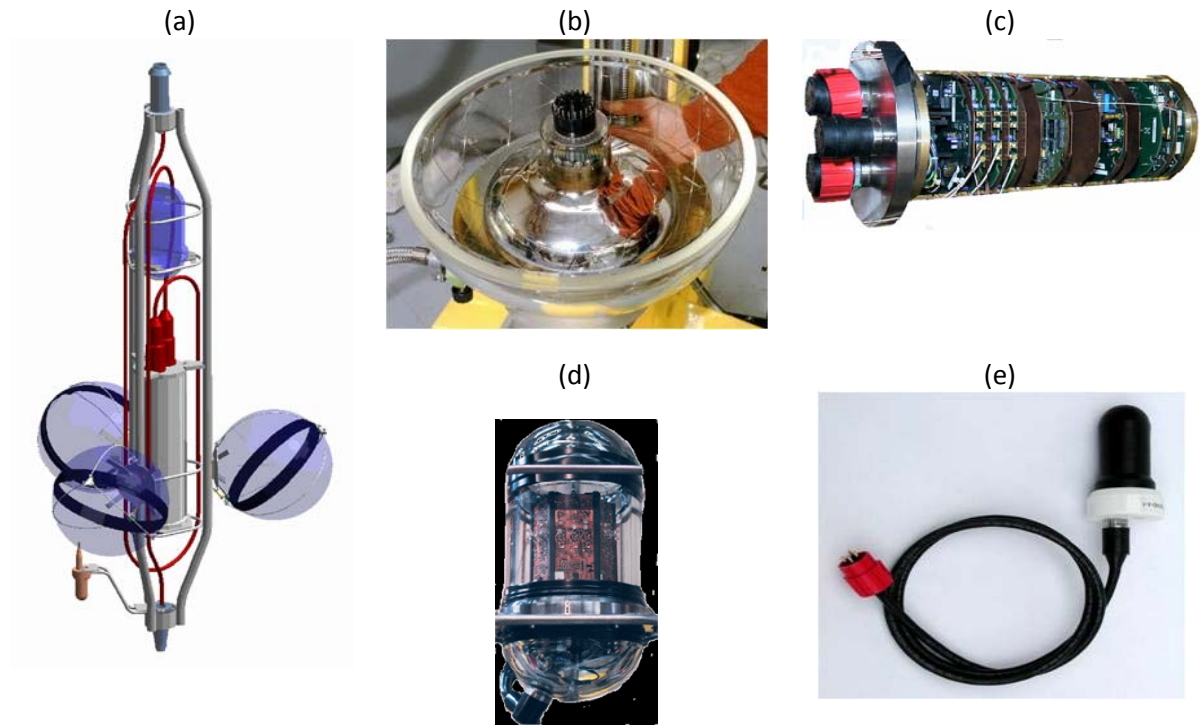

*Figure 1: The ANTARES storey (a), the phototube in a glass sphere, the optical gel and the mu-metal cage are visible (b), the electronics module (c). Some storeys contain a calibration flasher (d) and an acoustic receiver (e).*

For ANTARES the deployment of the first detector line, Line 1, occurred on February 14, 2006. The connection to the seabed communications cable occurred two weeks later using a submersible vehicle.

The detector line is situated at a depth of 2475 m approximately 40 km south east of the French town of Toulon. It consists of optical detectors attached to a cable which provides mechanical strength, electrical contact and fibre optic readout. This cable is held on the seabed by a deadweight anchor and kept vertical by a buoy at the top of its 450 m length.

Along this detector line 25 storeys of optical detectors are placed at an inter-storey distance of 14.5 m starting 100 m above the seabed. On each storey three spherical glass pressure vessels [**7**] contain 10" Hamamatsu photomultipliers (PMTs), which are oriented with their axes pointing downward at an angle of 45° from the vertical (see Figure 1).  A titanium cylinder houses the electronics for data acquisition and slow control. The orientation of each storey is determined continuously with the aid of a compass and a tilt meter. The position of every fifth storey is monitored using a hydrophone for acoustic triangulation. Front-end chips (analogue ring samplers - ARS) [**8**] sample the data from the photomultiplier tubes. If the pulse height exceeds a preset threshold the chips digitize the arrival time and charge of the PMT pulse. In order to limit dead time due to digitisation, each photomultiplier is sampled alternately by two ARS. For the analysis described in this paper the thresholds were effectively at the level of about 0.5 times the most probable pulse height produced by a single photoelectron (spe). Time and charge data of all hits are stored in memory. A master clock synchronizes the time stamping of this data. The time offsets between different PMTs are measured before deployment in a dedicated setup and then validated and monitored in situ by a system of calibration flashers distributed throughout the line [**9**]. In order to provide an efficient transfer of the data, the memory is divided into buffers of 104 ms length (a time slice). The starting times of the buffers on each storey are synchronised. These buffers are transmitted to shore via an Ethernet network to a farm of processors, where candidate events are filtered (online filter) from the continuous data stream for further processing offline. Individual processors in the farm receive and process all buffers from the full detector associated with a single time slice [**10**].

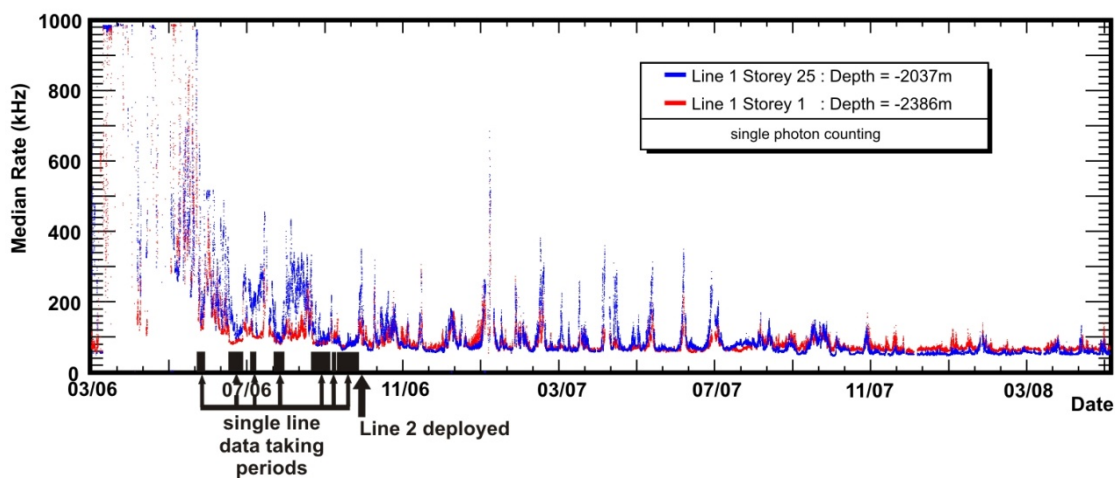

*Figure 2: The singles rate measured with a single OM, on both the lowest and highest floor of Line1, for the period March 2006 until the end of April 2008. The data taking periods for this analysis are*

*indicated.*

## 2. Data Selection

Since the start of data taking the singles rate (i.e. the single photon count rate) from each optical module has been monitored. Figure 2 shows the singles rate observed on Line 1 from March 2006 until the end of April 2008. At the beginning of the data taking period the singles rates increased from the expected 50-100 kHz per PMT to over 250 kHz, with rates exceeding 1 MHz in some periods. This increase in rate was due to increased bioluminescent activity. The onset of this high background period coincided with a period of unusually high water currents. While normal current values are typically below 5 cm/s the onset of this high background period was characterised by current speeds of up to 35 cm/s. The cause of these effects is still under study. After three months, however, the noise rates subsided and returned to the anticipated levels of 50-100 kHz. The expectation from the presence of the radioactive isotope $^{40}$K in the seawater is a singles rate of around 50 kHz. Since the end of June 2006 the fraction of time that the rate has surpassed 250 kHz is less than 5%. The high rate period was challenging for both data acquisition and filtering.

On the 26$^{th}$ of September 2006 the second line of the detector was connected and the period of single line running was terminated. For this paper data runs were selected with the singles rate averaged over all active components less than 120 kHz. In addition to the steady singles rate some optical activity arrives in short bursts, most likely associated with bioluminescence emitted by macroscopic organisms. To avoid periods where this burst activity was too large we also required that the fraction of time the singles rate was above 250 kHz should be less than 0.15.

The periods passing these criteria in the six months of exclusively one line data taking are indicated in Figure 2. The integrated acquisition time for the one line setup was 448 hours. In this period, however, the data acquisition was not yet fully functional and only every second time slice was transferred from the detector to shore. In addition an effective dead-time of 20% was present in the online data filtering process. These issues have since been corrected for later data taking periods. The total effective live time of the experiment was therefore 187 hours.

Events were filtered from the data stream by searching within the time slice of 104 ms for PMT signals that were correlated in position and time. Seeds for the search, level 1 trigger hits (L1 hits), were either coincidences between PMTs within a single storey, in a 20 ns time window, or single PMTs with a pulse height in excess of 10 spe. If a cluster of at least five L1 hits was found, where for any pair $|\Delta t| < \frac{n}{c}|\Delta \vec{x}| + 20$ ns, a candidate event was considered to be found. Here $|\Delta t|$ is the absolute time difference between the hits, $|\Delta \vec{x}|$ is the distance in space between the PMT pairs involved, $n = 1.34$ is the refractive index of the sea water and c is the velocity of light. All information from the full detector in a time window, the snapshot, extending from 2 µs before the first L1 to 2 µs after the last L1 in the cluster was stored for further offline processing. In total $7.5 \times 10^4$ candidates were stored.

## 3. Reconstruction

In this analysis the detector is assumed to be a single perfectly straight vertical line. All optical modules are taken to be stationary at the same position $x = y = 0$ ($x$ and $y$ are the horizontal coordinates, $z$ the vertical). As their orientation is not used in the analysis there is no information from which the azimuthal angle of the track can be determined. In reality the line is subject to forces due to the undersea water currents. Two factors will then contribute to the uncertainty in the position of the optical modules: the rotation of a storey around its axis and the deformation of the flexible line. The distance from the centre of the photocathode of the PMT to the axis is 57.9 cm. Ignoring the rotation of a storey introduces an uncertainty of 40 cm. This error on the position corresponds to a timing inaccuracy of about 2.0 ns. Deformation of the line shape introduces relative offsets in the $x - y$ plane between storeys. As a result, an error in the relative positioning of the storeys and thus of the optical modules is introduced. In the selected data taking period the water velocities were such that 94% of the time the top of the line did not deviate horizontally more than one metre. When the detector line is tilted, while assuming a vertical orientation, the true zenith is displaced with respect to the assumed zenith. A displacement of the top of the line of one metre corresponds to a shift of the zenith of about 0.15°. All results presented in this paper include these orientation effects and where relevant they are included in the systematic errors.

The track model [**11**] used in the reconstruction is that of a relativistic muon traversing the detector and emitting Cherenkov light. The fit is performed following the assumption of a single muon track, although a large fraction of the events is composed of muon bundles. The arrival time $t_j$ of Cherenkov photons at the PMT $j$ located on the line at vertical height $z_j$ is then defined by

$$c(t_j - t_d) = (z_j - z_d)\cos\theta + \sqrt{n^2 + 1}\sqrt{d^2 + (z_j - z_d)^2 \sin^2\theta}, \qquad (3\text{-}1)$$

with $d$ being the track's distance of closest approach to the line, $z_d$ and $t_d$ being the height and time at which the closest distance is reached, $\theta$ being the zenith. These variables are shown in Figure 3.

The L1 hits that caused the trigger are used as a starting point for the reconstruction. For L1s, which are comprised of coincident hits, only the earliest time is considered. Due to the non-linear nature of the fitting procedure, a series of fits is performed varying the track direction, $\theta$. Given this direction the muon trajectory is determined by three remaining parameters, which fix the spatial and temporal position of the track. These parameters are $t_0$, the time at which the track intersects a plane perpendicular to the assumed track direction and passing through the centre of the active part of the line and $x_0$ and $y_0$, at which the track crosses this plane (see Figure 3). All positions of the hit PMTs are projected onto this plane. The time at which PMT $j$ is hit is given by $t_j = t_o + \frac{1}{c}(z'_j + \kappa r_j)$ where $z'_j$ is the signed distance of PMT $j$ to the plane (positive in the direction of travel of the muon), $r_j$ is the distance of PMT $j$ to the track and $\kappa = \frac{c}{v_g}\frac{1}{\sin\theta_c} - \frac{1}{\tan\theta_c}$ with $v_g$ being the group velocity of light in water and $\theta_C$ being the Cherenkov angle. Using these definitions a criterion can be defined which is more restrictive than the causality criterion used in the online filter:

$$[-\kappa R + z'_j - z'_i]/c - 20\ ns \leq t_j - t_i \leq [\kappa R + z'_j - z'_i]/c + 20\ ns \qquad (3\text{-}2)$$

where $R$ is the projected distance on the plane between PMT i and j and the 20 ns have been included as safety margin. In addition it is required that two correlated hits not be further apart than 90 m, two times the absorption length of light in water [**12**]. The L1 hits of the filter are subjected to these more stringent criteria and if a cluster of at least 5 L1 hits remain then additional single hits,

L0s, are sought. If an L0 obeys the criteria with respect to all hits in the cluster it is added to the cluster.

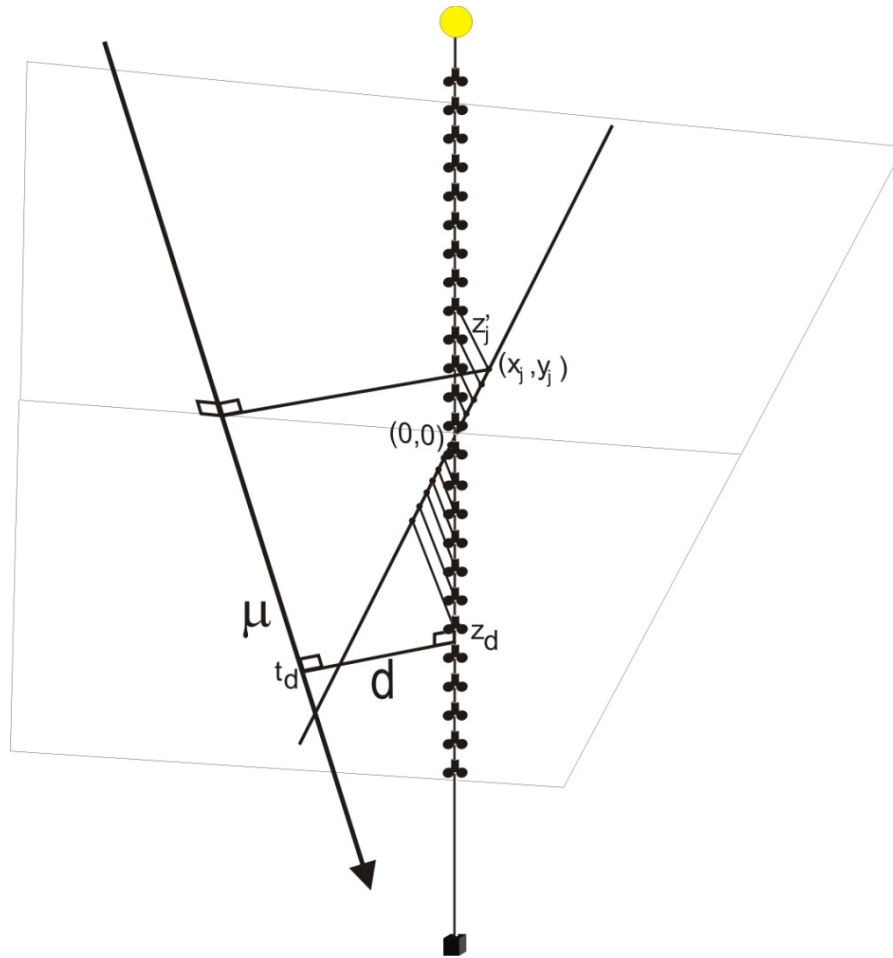

*Figure 3: The muon track intersecting the plane perpendicular to the track and passing through the centre (0,0) of the active part of the line. Parameters used in the tracking are also indicated.*

The hits in the cluster are then used to fit the three track parameters, $x_0$, $y_0$ and $t_0$. This is done in several steps. The starting values of the parameters are determined from the average position and time, calculated from all hits in the cluster weighted by their pulse height. First the hit residuals are minimised using Powell's method and this is then followed by Marquardt's method of $\chi^2$ minimization [**13**]. This full procedure is repeated for different values of the zenith angle, between 0° and 180°, in steps of 5°. If a sufficiently large cluster is found each separate angle will give an optimally fitted track and a set of hits used in the fit. For each fit a new $\chi^2$ minimization is performed with θ now left free as an additional parameter. After convergence all fits with a $\chi^2$ probability $P_{\chi^2} > 0.01$ are kept. For fits with $P_{\chi^2} \leq 0.01$, hits with an individual contribution to the $\chi^2$ of more than 25 are removed one by one and the fit is redone and $P_{\chi^2}$ recalculated. If three hits are removed and $P_{\chi^2}$ remains below 0.01, the fit is rejected.

All fits are finally ordered according to the number of hits used in the fit. If multiple solutions have the same number of hits, the one with the smallest $\chi^2$ is considered the best. Solutions with less than 20% of the hits of the best solution are rejected.

Two events are displayed in Figure 4 and Figure 5. In these figures the y-axis corresponds to the vertical height along the line and the x-axis displays the arrival time of the photons. The event shown in Figure 4 has been reconstructed as a muon passing the detector line in the vertically downward direction. All storeys of the line have been used in determining the fit. The hit times can be seen to propagate down the line at the speed of light, following the relativistic muon's Cherenkov cone propagation.

The event shown in Figure 5 shows a more complex structure, nevertheless a single downward-going track has been fitted. Approximately at the centre of the line a parabolic pattern of hits is superimposed on the track hits. These are most likely due to a bremsstrahlung photon having been radiated from the downward-going muon, causing an electromagnetic cascade. The Cherenkov light produced by this cascade is detected on the line as the equivalent of a point source emitting light. If these hits are used in the track fit an alternative, approximately horizontal, track solution is found. Both solutions are indicated in the figure. In this event the best solution (most hits) is the downward-going track; in other cases the horizontal track can turn out to be the best solution.

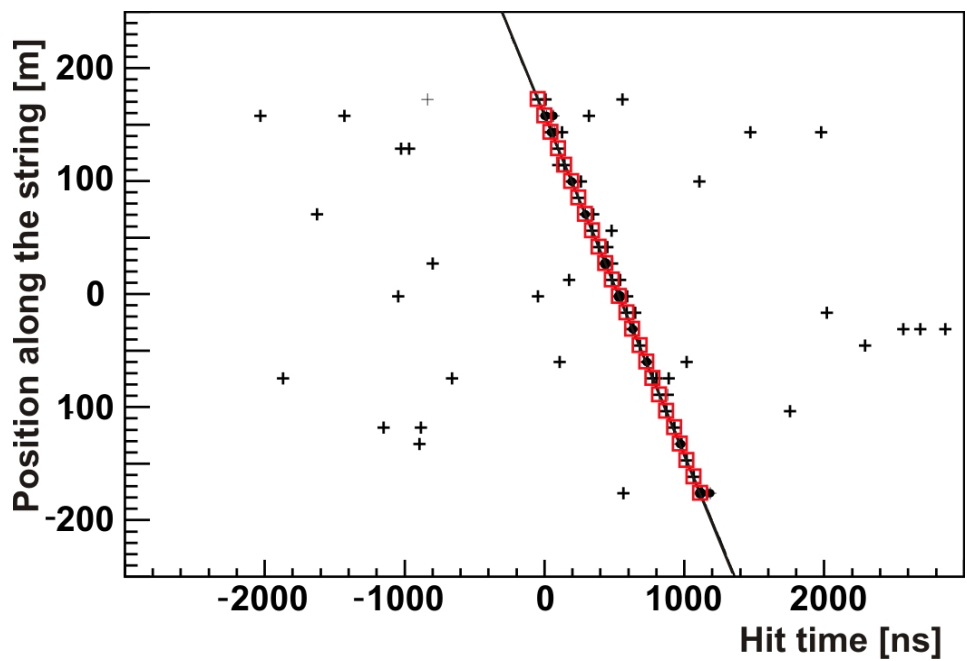

*Figure 4: A muon traversing the detector in the downward vertical direction. Crosses indicate hits, closed circles are the hits selected by the online filter algorithm. Hits used in the fit are indicated by open squares. For this particular track all storeys participate in the fit.*

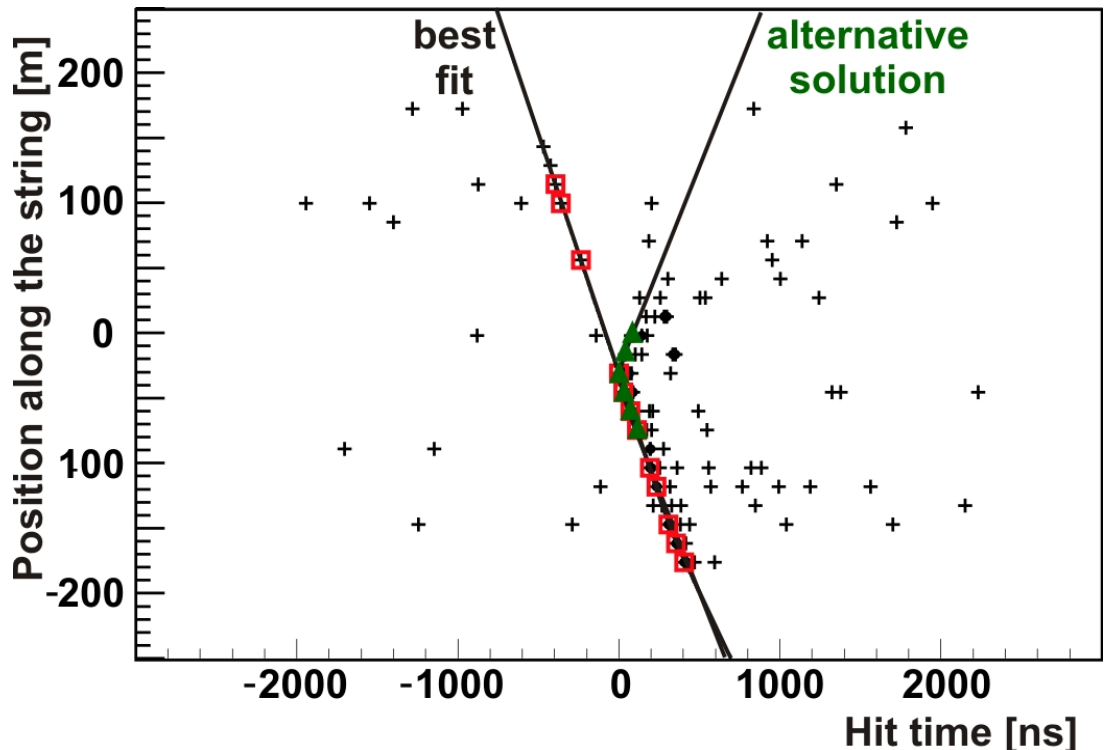

*Figure 5: A downward going muon with two possible track solutions. The best solution is a vertical track. The alternative solution is a horizontal track. Open squares indicate the hits used in the best fit, closed triangles those used in the alternative solution.*

## 4. Results

The distribution of time residuals of all hits in the snapshot with respect to the predictions from the track parameters, obtained from the best fit, is shown in Figure 6. The fit is performed according to the procedure described above but excluding the hit under consideration. This is done to obtain an unbiased measure of the residual. There is a sharp peak at zero time difference, corresponding to photons arriving at the time expected from the track model. A broad enhancement is seen starting at 45 ns. This is due to the loss of the time information of a second pulse from the PMT, which arrives either before the end of the 25 ns integration gate or during the 15 ns dead time of swapping from one ARS to the other. The long tail towards late arrival times is due either to light emitted from bremsstrahlung showers or to scattered light. The small shoulder occurring at 20 ns is due to the trigger requirement. Since each fit contains five L1s, and since the fit uses only the earliest time of each L1 component-hit, the remaining component-hits give rise to the shoulder, which has a width of 20 ns and is equal to the L1 coincidence window.

Finally, the small constant level of early hits is due to random background, mostly from $^{40}$K, whereas the small shoulder at early times is due to poorly reconstructed tracks. The RMS width of the central peak depends on the number of hits used in the fit. As shown in Figure 7, it decreases from 12 ns for tracks with the minimum number of 5 hits, to 2 ns for the maximum number of 24 hits.

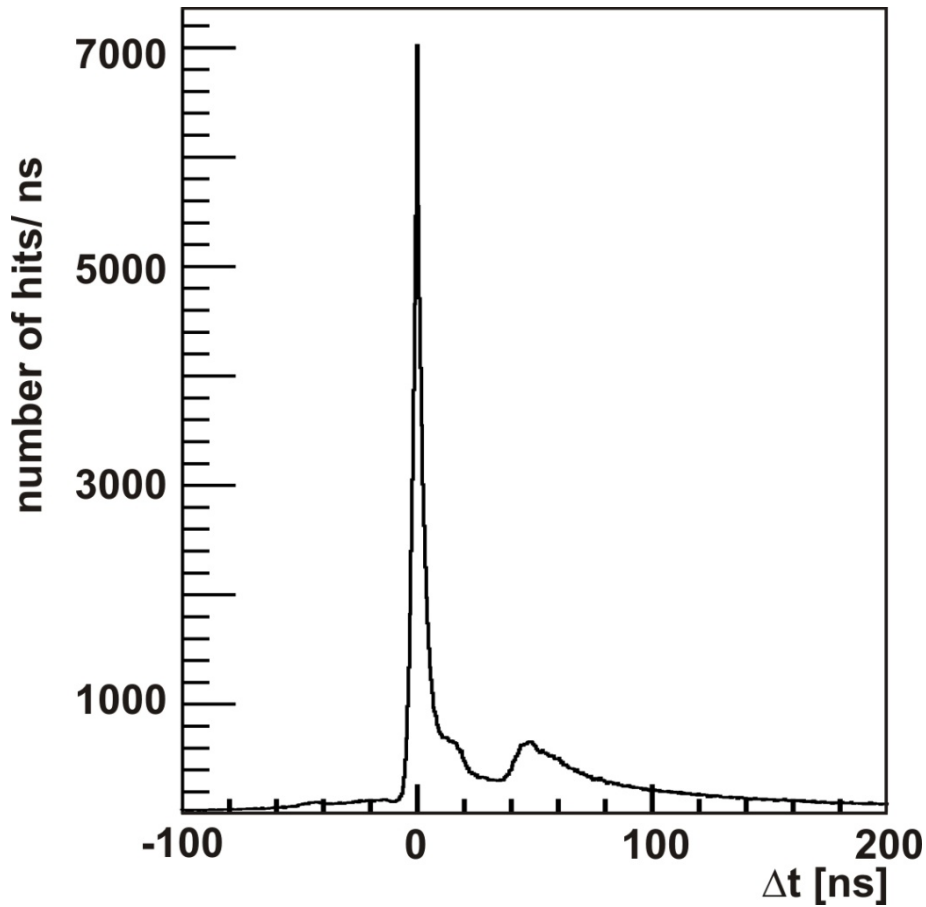

*Figure 6: The residual distribution for all hits in the snapshot. See the text for an explanation of the different structures observed in the distribution.*

To investigate the tracking method in more detail, a simulation of downward going muons from extended air showers has been performed. This simulation uses the parameterization of the muon flux, given by MUPAGE [14], at the depth of the ANTARES detector. All tracks that pass within two and a half light attenuation lengths, or 135 m, from the detector line are passed to the program KM3 [15] that performs the detailed simulation of the detector response. This program uses parameterizations of light intensities as a function of the distance from the light source that incorporate the scattering of light in water, the effects of light dispersion and a detailed simulation of the PMT and subsequent ARS response. Optical noise, in the form of random single photon hits with rates commensurate with those measured during the runs used for this analysis, is added to these simulated events. As can be seen from Figure 7, this simulation reproduces the time residuals very well. It also indicates that the degraded resolution at small numbers of hits is due to poor reconstruction, where hits originating from electromagnetic showers or from more than one muon are combined in a single track fit.

By taking the best track solution for each event an estimate of the zenith angular resolution has been obtained. First, all tracks were refitted using hits from even numbered floors only or from odd numbered floors only. These two solutions were then compared and the angular difference determined. The size of such angular difference depends on the number of hits used in the fit; it decreases from around $2^o$ for tracks with only five hits to below $1^o$ for tracks with more than eight hits. Since both tracks contribute equally, the angular resolution for each track is therefore around 0.7°. Taking into account the improvement to be expected when the track is fitted over its full length, the full zenith-angular resolution for tracks with more than 16 hits on only one detector line approaches 0.5°.

As the optical modules are facing downwards at 45°, the acceptance for Cherenkov light for downward going muons is limited. It depends quite strongly on the sensitivity of the optical module for light arriving from the rear. The sensitivity of the optical modules has been measured [7] and simulated using ray tracing through the different parts of the optical module. The results differ for photons arriving with an incident angle of more than 90° with respect to the axis of the PMT. For upward-going tracks the differences have a negligible effect on the acceptance. However, for downward-going muons, these differences are significant and lead to an increase in the systematic uncertainties. New measurements are being performed in order to better understand the angular acceptance of the optical modules.

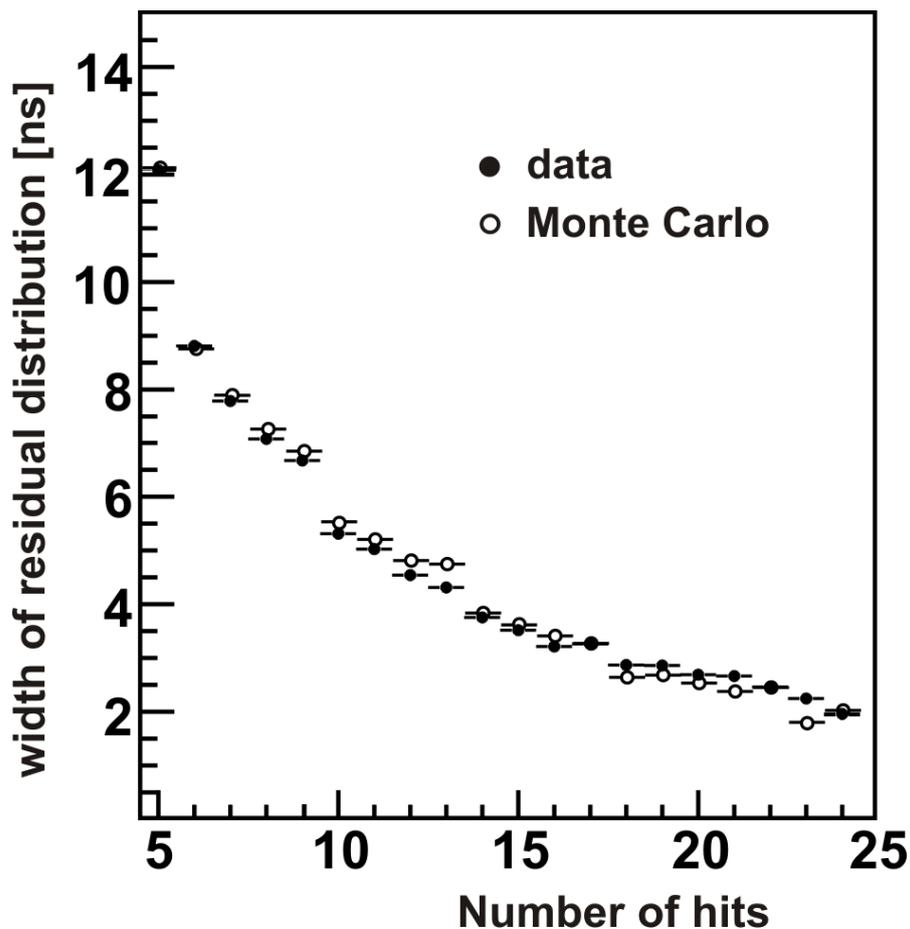

*Figure 7: The RMS width of the central peak of Figure 6 as a function of the number of hits used in the fit. Closed circles represent the data and open circles are the results of the fits to simulated data.*

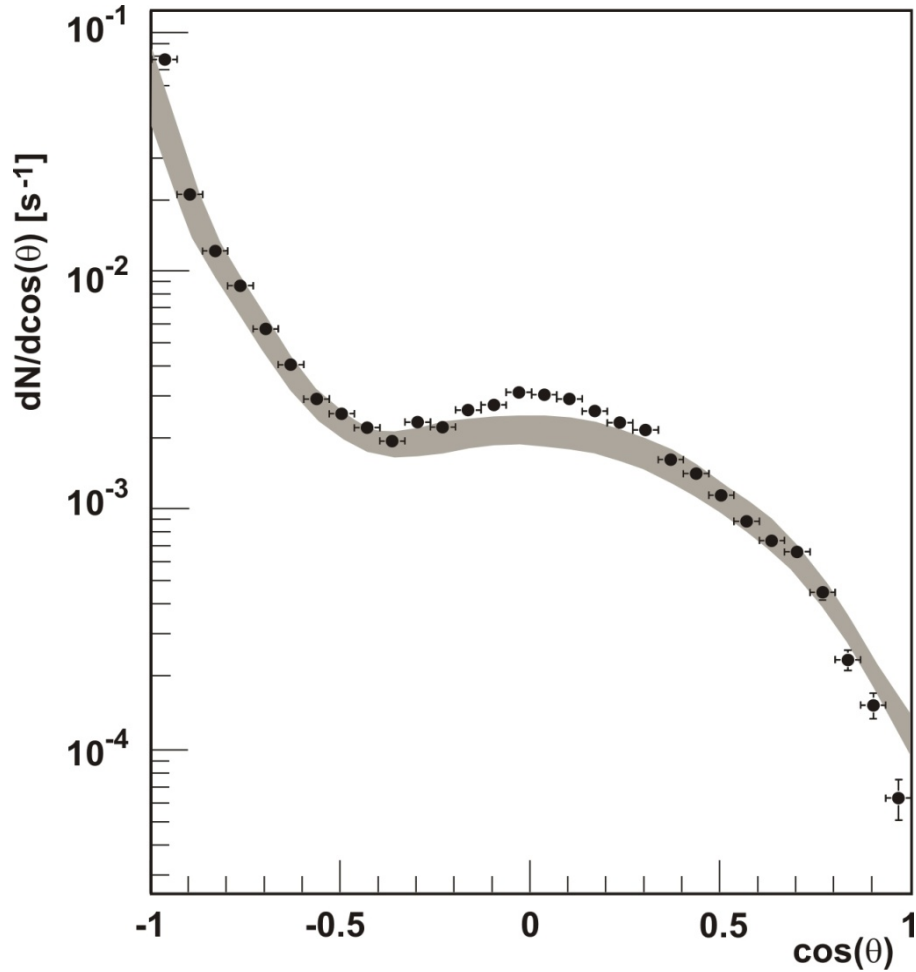

*Figure 8: Distribution of the cosine of the angle with respect to the vertical for all tracks extracted from the ANTARES Line 1 data. A value of -1 corresponds to downward-going muons. The points are data; the shaded band is the result of the simulation. Errors on the data points are statistical only. The width of the band is due mainly to the uncertainty in the angular acceptance of the PMT.*

The angular distribution of the best fit solutions found for all data obtained with Line 1 before October 2006 is shown in Figure 8. The distribution follows the expected pattern for small angles with respect to vertically downward-going muons ($\cos\theta = -1$). As the angle moves away from vertically down-going, the intensity decreases rapidly. At somewhat larger angles the distribution flattens out and rises. The simulation shows a similar behaviour, although not quite as strong. In the simulation this feature is caused by vertical tracks producing bremsstrahlung showers, which are mistakenly interpreted as approximately horizontal tracks. The size of this effect is sensitive to the angular distribution of photons emitted in bremsstrahlung showers. In the absence of information from additional detector lines, this misreconstruction is unavoidable. The simulations show that for values of $\cos\theta < -0.5$, there is minimal distortion of the angular distribution and the acceptance correction is flat in this region. The region with $\cos\theta > -0.5$ becomes more and more dominated by poorly reconstructed vertical tracks.

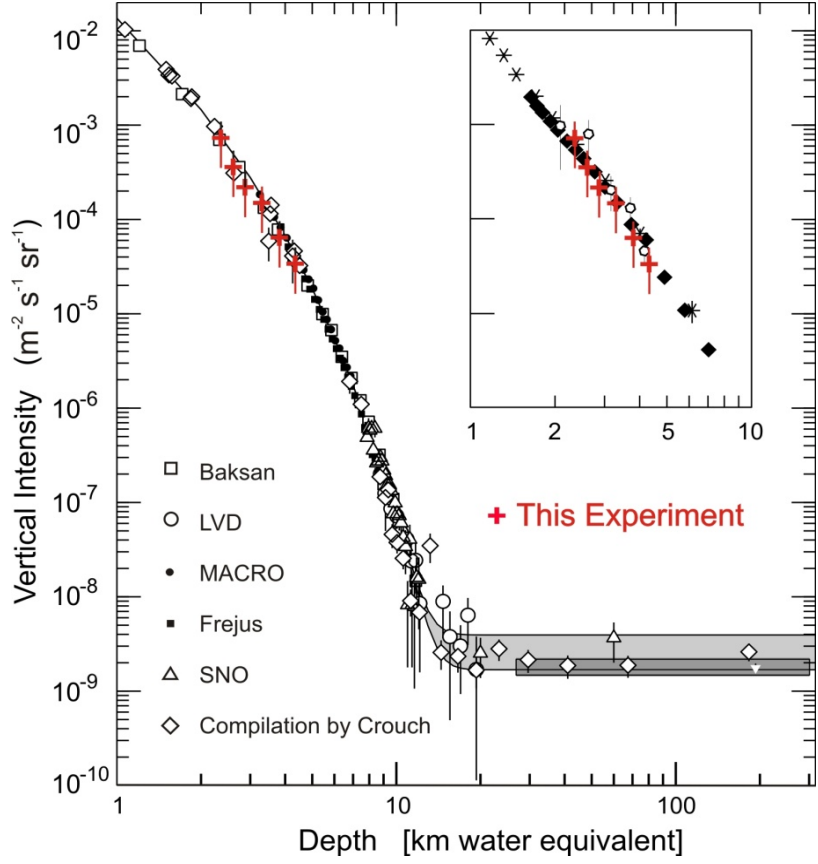

*Figure 9: Vertical muon intensity versus depth. Crosses are the results of the present experiment, other data are from: the compilations of Crouch [16], Baksan [17], LVD [18], MACRO [19], Frejus [20], and SNO [21]. The shaded area at large depths represents neutrino-induced muons of energy above 2 GeV. The upper line is for horizontal neutrino-induced muons, the lower one for vertically upward muons. In the inset the data from the present experiment is compared to other water or ice experiments [22,23,24]. For the present experiment the measurements correspond to 187 hours of data taking.*

To measure the vertical muon intensity versus depth, the measured rate of reconstructed tracks is converted to the single muon intensity using the Monte Carlo simulation. Only tracks having $\cos\theta < -0.5$, where the conversion is almost independent of angle, is used. Each value of the zenith angle corresponds to a slant depth through the water mass above the detector. Therefore the measured single muon flux can also be given as a function of slant depth. In order to calculate the vertical muon intensity, the energy distribution of muons at sea level has to be taken into account. This distribution for small angles, i.e. $\pi - \theta$ small, is given by [25]:

$$\frac{dN_\mu}{dE_\mu} \approx 0.14\, E_\mu^{-2.7} \left( \frac{1}{1 - \frac{1.1 E_\mu \cos\theta}{115}} + \frac{0.054}{1 - \frac{1.1 E_\mu \cos\theta}{850}} \right) [\text{cm}^{-2}\text{s}^{-1}\text{sr}^{-1}\text{GeV}^{-1}] \qquad (4\text{-}1)$$

All energies are given in GeV. For a given value of $\cos\theta$ this distribution is integrated over the energy range from 2 TeV to 10 PeV to yield $R(\cos\theta)$. The correction factor $F(\cos\theta) = R(\cos\theta)/R(-1)$ is used to correct the intensity as a function of slant depth to the vertical muon intensity. The measured values of the vertical muon intensity versus depth vary between $(7.4 \pm 3.5) \times 10^{-4}$ $m^{-2}s^{-1}\,sr^{-1}$ at 2400 m to $(3.8 \pm 1.9) \times 10^{-5}\ m^{-2}s^{-1}\,sr^{-1}$ at 4800 m. The data are presented in Figure 9 and compared to other measurements compiled by the Particle Data Group [26]. The errors are systematic and of the order of 50%. They are dominated by the uncertainty in the PMT acceptance, with variations observed in the PMT thresholds yielding an error of 6% and variations in the filter dead-time giving a further 5% error. All other errors including the statistical errors are negligible. Agreement between the present data and other published values is good. Our data are also consistent with [5].

## 5. Conclusions

We have presented data on the performance of the first detector line deployed by the ANTARES collaboration. Although bioluminescent activity was extreme in the first months after deployment such activity has subsided. No technical problems have been encountered during the two years of running. The full readout and reconstruction chain has been described. The measured angular distribution of the reconstructed tracks agrees reasonably well with the results from simulations of muons from extensive air showers. The major systematic error on the acceptance of these mainly downward-going tracks is due to the uncertainty in the angular acceptance of the PMT, especially in the rearward direction. The measured intensity as a function of depth of these muons agrees well with previous measurements. The zenith angular resolution of tracks with 16 or more hits on a single line is found to be of the order of half a degree.

## 6. Acknowledgements

The authors acknowledge the financial support of the funding agencies: Centre National de la Recherche Scientifique (CNRS), Commissariat à l'Energie Atomique (CEA), Commission Européenne (FEDER fund and Marie Curie Program), Région Alsace (contrat CPER), Région Provence-Alpes-Côte d'Azur, Département du Var and Ville de La Seyne-sur-Mer, in France; Bundesministerium für Bildung und Forschung (BMBF), in Germany; Istituto Nazionale di Fisica Nucleare (INFN), in Italy; Stichting voor Fundamenteel Onderzoek der Materie (FOM), Nederlandse organisatie voor Wetenschappelijk Onderzoek (NWO), in The Netherlands; Russian Foundation for Basic Research (RFBR), in Russia; National Authority for Scientific Research (ANCS) in Romania; Ministerio de Ciencia e Innovación (MICINN), in Spain.
We also acknowledge the technical support of Ifremer, AIM and Foslev Marine for the sea operation and the CC-IN2P3 for the computing facilities.